\documentclass[prd,twocolumn,superscriptaddress,preprintnumbers,balancelastpage,nofootinbib,floatfix]{revtex4-2}

\usepackage{graphicx}  
\usepackage{dcolumn}   
\usepackage{bm}        
\usepackage{amssymb, amsmath}
\usepackage{slashed}   
\usepackage{lipsum}
\usepackage[usenames,dvipsnames,svgnames,table]{xcolor}
\usepackage{comment}
\usepackage[utf8]{inputenc}
\usepackage{multirow}
\usepackage{subfigure}
\usepackage{xfrac} 

\definecolor{BlueViolet}{rgb}{0.2, 0.00, 0.7}
\definecolor{Blue}{rgb}{0.15, 0.00, 0.9}
\definecolor{lightblue}{rgb}{0.15, 0.35, 0.95}
\definecolor{kitgreen}{rgb}{0,
0.58823 
, 0.50980 
}
\usepackage[
colorlinks=true, linkcolor=lightblue,citecolor=lightblue,urlcolor=kitgreen]{hyperref} 

\newcommand{\ov}{\overline}

\newcommand{\bea}{\begin{eqnarray}}
\newcommand{\eea}{\end{eqnarray}}

\newcommand{\bbms}{\bbs\ mixing}

\newcommand{\bbs}{\ensuremath{B_s\!-\!\Bbar{}_s\,}}

\newcommand{\TeV}{\,\text{TeV}}

\newcommand{\eg}{{\em e.g.}}
\newcommand{\ie}{{\em i.e.}}
\usepackage{adjustbox}
\usepackage{tabularx}
\newcolumntype{Y}{>{\centering\arraybackslash}X} 
\usepackage{booktabs} 
\usepackage{colortbl} 
\def\beq#1\eeq{\begin{align}#1\end{align}}

\RequirePackage{xspace}
\def\Bbar    {\kern 0.18em\overline{\kern -0.18em B}{}\xspace}

\begin{document}
\preprint{ZU-TH 32/25, KEK-TH-2718, CHIBA-EP-271}

\title{Discriminating Tauphilic Leptoquark Explanations of the \texorpdfstring{\boldmath{$B$}}{B} Anomalies\\[0.2cm]
via \texorpdfstring{\boldmath{$K\to \pi\nu\ov\nu$}}{K to pi nu nubar} and \texorpdfstring{\boldmath{$B\to K\nu\ov\nu$}} {B to K nu nubar}}

\author{Andreas Crivellin}
\email{andreas.crivellin@cern.ch}
\affiliation{Physik-Institut, Universit\"{a}t Z\"{u}rich, Winterthurerstrasse 190, CH–8057 Z\"{u}rich, Switzerland}

\author{Syuhei Iguro}
\email{igurosyuhei@gmail.com}
\affiliation{Institute for Advanced Research, Nagoya University, Nagoya 464-8601, Japan}
\affiliation{Kobayashi-Maskawa Institute for the Origin of Particles and the Universe, Nagoya University, Nagoya 464-8602, Japan}
\affiliation{KEK Theory Center, IPNS, KEK, Tsukuba 305-0801, Japan}

\author{Teppei Kitahara}
\email{kitahara@chiba-u.jp}
\affiliation{Department of Physics, Graduate School of Science,
Chiba University, Chiba 263-8522, Japan}
\affiliation{Kobayashi-Maskawa Institute for the Origin of Particles and the Universe, Nagoya University, Nagoya 464-8602, Japan}

\begin{abstract}
Leptoquark models are prime candidates for new physics (NP) explanations of the long-standing anomalies in semi-leptonic $B$ decays;  $b\to c \tau \ov\nu$ (encoded in $R(D^{(\ast)})$) and $b\to s\ell\ov\ell \,(\ell=e,\mu)$ transitions. Furthermore, Belle~II and NA62 reported weaker-than-expected limits on $B^+ \to K^+ \nu\ov\nu$ and  $K^+ \to \pi^+ \nu \ov\nu$, respectively.
While the $R(D^{(\ast)})$ and $b\to s\ell \ov\ell$ measurements can be explained with NP contributions at the $\mathcal{O}(10\%)$ level, the neutrino channels suggest that the NP effect could be comparable in size to the Standard Model one. In this context, we consider the two types of leptoquark models with minimal sets of the couplings that can best describe the semi-leptonic $B$ anomalies and lead at the same time to effects in the neutrino modes, the singlet-triplet scalar leptoquark model ($S_1+S_3$) and the singlet vector leptoquark model ($U_1$). More specifically, the neutrino channels pose non-trivial constraints on the parameter space, and we find that large effects (\ie, accounting for the current central value) in $B\to K^{(*)}\nu\ov\nu$ are only possible in the $S_1+S_3$ setup, while both models can account for the central value of $K^+\to \pi^+\nu\ov\nu$.
\\
\end{abstract}

\maketitle

\section{Introduction}

Precision measurements have the potential to reveal the effects of physics beyond the Standard Model (SM). In particular, the study of kaon decays has been leading this frontier and can, in some instances, push the new physics (NP) mass scale to above $10^4$\,TeV~\cite{Cerri:2018ypt,Buras:2024mhy}. Among these measurements, $s\to d \nu \ov{\nu}$ transitions are promising to reveal NP effects since they are dominated by calculable short-distance contributions, which are Glashow-Iliopoulos-Maiani (GIM) suppressed~\cite{Glashow:1970gm} and involve small Cabibbo-Kobayashi-Maskawa (CKM) elements~\cite{Kobayashi:1973fv}, thus leading to quite small branching ratios of $\mathcal{O}(10^{-11})$ with controlled theoretical uncertainties~\cite{Buras:2023qaf}. Furthermore, the corresponding measurements have achieved sensitivities of $\mathcal{O}(10^{-10})$, thus exceeding the precision of $B$ physics by orders of magnitude. 

The latest NA62 result for $K^+ \to \pi^+ \nu\ov\nu$~\cite{NA62talk,NA62:2024pjp}
\begin{eqnarray}
    {\mathcal{B}}(K^+\to \pi^+ \nu \ov{\nu})= \left(13.0^{+3.3}_{-3.0} \right)\times 10^{-11}\,,
\end{eqnarray}
results in a more than $5\sigma$ rejection of the background-only hypothesis. This first observation of $K^+\to \pi^+ \nu \ov{\nu}$ should be compared with the SM prediction~\cite{Buras:2015qea,Brod:2021hsj,Buras:2022wpw,DAmbrosio:2022kvb,Allwicher:2024ncl}\footnote{The numerical values of the CKM elements used for this prediction are given in Sec.~\ref{sec:pheno}.}
\begin{eqnarray}
    {\mathcal{B}}(K^+\to \pi^+ \nu \ov{\nu})_{\rm{SM}}= \left(8.09 \pm 0.63 \right)\times 10^{-11}\,.
\end{eqnarray}
From this, we obtain the observed signal strength with respect to the SM of\footnote{For the neutral kaon mode, the upper limit $\mathcal{B}(K_L \to \pi^0 \nu \ov\nu) < 2.2 \times 10^{-9}$, obtained by the KOTO experiment~\cite{KOTO:2024zbl}, should be compared with the SM prediction $\mathcal{B}(K_L \to \pi^0 \nu \ov\nu)_{\text{SM}}= \left( 2.58 \pm 0.30 \right)\times 10^{-11}$~\cite{Allwicher:2024ncl}. While this currently does not lead to competitive bounds, an improvement by three orders of magnitude is foreseen by the proposed KOTO~II experiment~\cite{KOTO:2025gvq}.}
\begin{align}
\begin{aligned}
    \mu (K^+ \to \pi^+ \nu\ov\nu)
   & \equiv \frac{\mathcal{B}(K^+\to \pi^+ \nu \ov{\nu})}{\mathcal{B}(K^+\to \pi^+ \nu \ov{\nu})_{\rm{SM}}}=1.61^{+0.43}_{-0.39}\,.
\end{aligned}
\end{align}

Similar to $K^+ \to \pi^+  \nu\ov\nu$, an $\mathcal{O}(1)$ NP effect is preferred by $B \to K^{(\ast)} \nu \ov\nu$ data. More specifically, the Belle~II collaboration announced an excess in their $B^+ \to K^+ \nu\ov\nu$ search~\cite{Belle-II:2023esi}, which, combined with the Belle~\cite{Belle:2013tnz, Belle:2017oht} and BaBar measurements~\cite{BaBar:2010oqg,BaBar:2013npw}, 
results in a world average~\cite{Belle-II:2023esi} of
\begin{eqnarray}
    {\mathcal{B}}(B^+\to K^+ \nu \ov{\nu})= \left(1.3\pm 0.4\right)\times 10^{-5}\,.
\end{eqnarray}
This is three times larger than the SM value~\cite{Becirevic:2023aov}
\begin{eqnarray}
    {\mathcal{B}}(B^+\to K^+ \nu \ov{\nu})_{\rm{SM}}= \left(0.472\pm0.027 \right)\times 10^{-5}\,.
\end{eqnarray}
They lead to a signal strength\footnote{%
The upper limit ${\mathcal{B}}(B^0\to K^{0} \nu \ov{\nu})<2.6\times 10^{-5}$, which has been set by the Belle~\cite{Belle:2017oht},
is much weaker than Eq.~\eqref{eq:ssBKnunu}.} 
\begin{align}
\mu (B^+\to K^+ \nu \ov{\nu}) = 2.75 \pm 0.86\,,
\label{eq:ssBKnunu}
\end{align}
which is consistent with the SM ($\mu=1$) at $\approx2\sigma$. Note that for $B\to K^{*} \nu \ov{\nu}$, only an upper limit $\mathcal{B}(B\to K^{*} \nu \ov{\nu})<2.7 \times 10^{-5}$ has been set by Belle~\cite{Belle:2017oht} while the corresponding Belle~II result is still pending.\footnote{%
Relative to the SM prediction an upper limit $\mu(B\to K^{*} \nu \ov{\nu}) \lesssim 2.7$ is obtained. This limit is not in conflict with Eq.~\eqref{eq:ssBKnunu}. Note that NP models studied in this paper have a purely left-handed current, so the relative effect in all $b\to s\nu\ov\nu$ transitions is the same. However, since the relative efficiency in $B\to K^{*} \nu \ov{\nu}$ vs $B\to K \nu \ov{\nu}$ is not given in the experimental articles, we do not combine both modes our phenomenological analysis but rather just show the preferred regions from the latter and the exclusion limit from the former.}

This situation in decays with neutrino final states is particularly interesting in light of the (possibly) related long-standing anomalies in the semi-leptonic $B$-meson decays~\cite{Capdevila:2023yhq}.\footnote{See, \eg,~Refs.~\cite{Fajfer:2018bfj,Descotes-Genon:2020buf,Allwicher:2023xba,Chen:2023wpb,Rosauro-Alcaraz:2024mvx,Chen:2025npb} for articles studying the correlations between the neutrino modes and the $B$ anomalies.} Already in 2012, the measurements of the lepton-flavor universality (LFU) in
\begin{align} 
R(D^{(\ast)})\equiv \frac{{\mathcal{B}}(B\to D^{(\ast)}\tau\ov\nu)}{{\mathcal{B}}(B\to D^{(\ast)} \ell\ov\nu)} \qquad (\ell= e,\mu)\,,
\end{align}
deviated from the SM prediction. The current HFLAV world averages~\cite{HFLAV:2022pwe}
\begin{align}
    \begin{aligned}
    R(D) &= 0.342 \pm 0.026\,, \\
    R({D^\ast}) & = 0.287 \pm 0.012\,.
    \label{eq:WAvalues}
    \end{aligned}
\end{align}
are still $\approx  15\%$ larger than the SM expectations~\cite{Bordone:2019vic,Bordone:2019guc,Iguro:2020cpg,Bernlochner:2022ywh}, which leads to $\approx  4\sigma$ deviation~\cite{Iguro:2024hyk}. Furthermore, very recently, these tendencies were confirmed by the latest Belle II measurement \cite{BelleIIMoriond,Belle-II:2025yjp}. Combining this measurement with the HFLAV value we obtain $\{3.8 \sigma, 4.7\sigma, 4.9 \sigma\}$ deviations, corresponding to $p\text{-value}=\{1.5 \times 10^{-4}, 2.4\times 10^{-6}, 1.2\times 10^{-6}\}$ ($\Delta \chi^2 = \{17.7, 25.9, 27.3$\} with two degrees of freedom) from the SM predictions of \{HFLAV2024~\cite{HFLAV:2022pwe}, Bernlochner~{\it et al.}~\cite{Bernlochner:2022ywh}, Iguro-Watanabe~\cite{Iguro:2020cpg}\}, respectively.

Furthermore, in neutral-current semi-leptonic $B$ decays, the LHCb collaboration observed already in 2013~\cite{LHCb:2013ghj,Aaij:2015oid} a tension in the angular observable $P_5^\prime$~\cite{Descotes-Genon:2013vna}, 
which still persists and was even recently confirmed by the CMS collaboration~\cite{CMS:2024atz}. This is accompanied by deficits in the total branching ratios of $\mathcal{B}(B\to K\mu\ov\mu)$~\cite{LHCb:2014cxe,LHCb:2016ykl,Parrott:2022zte} and $\mathcal{B}(B_s\to \phi \mu\ov\mu)$ (together with angular observables in the latter)~\cite{LHCb:2021zwz,Parrott:2022rgu,Gubernari:2022hxn} and semi-inclusive observables~\cite{Isidori:2023unk}. In conjunction with updated tests of lepton flavor universality, the $R(K^{(\ast)})$ ratios~\cite{LHCb:2022qnv} and the $Q$ observables~\cite{LHCb:2025pxz}, which all agree with the SM predictions, this indicates a lepton-flavor-universal (U) effect in $b\to s\ell \ov\ell$. 

In fact, the global fits to the Wilson coefficient $C_9^{\text{U}}$ of the effective Hamiltonian
\begin{align}
\begin{aligned}
\mathcal{H}_{\text{eff}}&=
- \frac{4 G_F}{\sqrt{2}} \frac{\alpha}{4\pi}V_{ts}^\ast V_{tb}C_{9}^\ell(\overline{s}\gamma_{\mu}P_{L} b)(\overline{\ell}\gamma^\mu \ell) 
+\rm{h.c.}\,,
\label{eq:C9U}
\end{aligned}
\end{align}
show statistically very significant indications for a non-zero new physics value~\cite{Alguero:2022wkd,Alguero:2023jeh,Hurth:2023jwr}.\footnote{%
Note that a sizable NP effect in $C_{10}^\mu$ is disfavored by the $B_s\to \mu\ov\mu$ measurement~\cite{ParticleDataGroup:2024cfk}.}

When attempting to construct a combined explanation of the charged and neutral current $B$ anomalies~\cite{Calibbi:2015kma,Buttazzo:2017ixm}, it is essential to keep in mind that to explain $R(D^{(\ast)})$, a tree-level effect is necessary which must be related to the tau lepton (and neutrino)~\cite{Fedele:2022iib,Ray:2023xjn,Fedele:2023ewe}. Interestingly, this setup naturally leads to an LFU effect in $C_9^{\text{U}}$ via off-shell photon-penguin involving the tau-lepton loop~\cite{Bobeth:2014rda,Crivellin:2018yvo}. This mechanism providing a combined explanation can be achieved in the $SU(2)_L$ singlet-triplet scalar leptoquark model ($S_1$ and $S_3$)~\cite{Crivellin:2017zlb,Crivellin:2019dwb,Gherardi:2020qhc} and the $SU(2)_L$ singlet vector leptoquark model ($U_1$)~\cite{Barbieri:2016las,Calibbi:2017qbu,Bordone:2017bld,Blanke:2018sro}.\footnote{Note that a photon penguin induced loop effect can also generate $C_9^{\text{U}}$ in the generic 2HDM~\cite{Iguro:2018qzf,Crivellin:2019dun,Crivellin:2023sig} or di-quark models~\cite{Crivellin:2023saq}.}

As the mechanism works the same way for both models, the question of how to disentangle them arises. In this paper, we will focus on $s \to d\nu\ov\nu$ and $b\to s \nu\ov\nu$ transitions, which, as we will see, have significantly different predictions in the two models. In particular, the $S_1+S_3$ model requires a tuning between the $S_1$ and $S_3$ contributions $b\to s\nu\ov\nu$ for explaining $R(D^{(*)})$ while being compatible with the $B\to K^{(*)}\nu\ov\nu$ bounds.\footnote{In particular, these constraints rule out a full explanation of $R(D^{(*)})$ with the $S_1$ model as proposed in Ref.~\cite{Bauer:2015knc}.} In general, the cancellation between the $S_1$ and $S_3$ neutrino contributions is expected to be imperfect and can result in a significant effect in $K\to \pi\nu\ov\nu$. This is in stark contrast to the $U_1$ LQ model, where such an effect is only generated at the loop level (but unavoidable)~\cite{Crivellin:2018yvo}.

Therefore, in this paper, we update these two promising leptoquark explanations of the semi-leptonic $B$ anomalies, and investigate the correlations with $K\to\pi\nu\ov\nu$ and $B\to K\nu \ov\nu$. For this, the paper is organized as follows. The leptoquark models are briefly summarized in Sec.~\ref{sec:model}. In Sec.~\ref{sec:pheno}, we discuss the relevant observables and perform phenomenological analyses. Section~\ref{sec:conc} is devoted to conclusions and discussions.

\section{Leptoquark Models}
\label{sec:model}
 
As the Introduction outlines, we will study the two leptoquark models $S_1+S_3$ and $U_1$, which provide combined explanations of the $R(D^{(*)})$ and $b\to s\ell\ov{\ell}$ anomalies.\footnote{We do not consider the $R_2$ leptoquark model here. The reason is that even though it could as well explain $b\to s\ell\ov{\ell}$ via an off-shell photon penguin through a tau loop, the loop effects in $b\to s\nu\ov\nu$ are highly suppressed~\cite{Crivellin:2022mff}.} While $S_1$ and $S_3$ are scalars which transform under the SM gauge group $SU(3)_c \times SU(2)_L \times U(1)_Y$ as $(\bar{{\bf 3}}, {\bf 1}, \sfrac{1}{3})$ and $(\bar{{\bf 3}}, {\bf 3}, \sfrac{1}{3})$ representations, respectively, $U_1$ is a vector with $({\bf 3}, {\bf 1}, \sfrac{2}{3})$. In our study, we assume the leptoquarks to couple only to third-generation leptons (taus and tau neutrinos) and set all leptoquark masses to 2\,TeV for concreteness, which is compatible with the collider bounds on the pair production of third-generation leptoquarks~\cite{CMS:2023qdw,ATLAS:2024huc}.\footnote{The bounds from single resonant production and non-resonant searches will be discussed in Sec.~\ref{sec:pheno}.}

\subsection{Singlet-triplet scalar leptoquark model: \texorpdfstring{\boldmath{$S_1+S_3$}}{S1+S3}}
\label{sec:model_S1S3}

The $S_1+S_3$ model was proposed in Ref.~\cite{Crivellin:2017zlb} and its phenomenology was studied in Refs.~\cite{Marzocca:2018wcf,Crivellin:2019dwb,Gherardi:2020qhc,Bordone:2020lnb,DaRold:2020bib,Marzocca:2021miv,Bhaskar:2022vgk}. We consider the following Yukawa-like interaction~\cite{Crivellin:2017zlb,Crivellin:2019dwb,Gherardi:2020qhc}
\begin{align}
{\mathcal{L}}_{\rm{S_1+S_3}}&=\left(\lambda^{L}_{ij}\ov{Q^C_i} \left(i\sigma^2\right)   L_j+\lambda^{R}_{ij}\ov{u^C}_i  e_j \right) S_1 \notag\\
    &\quad +\kappa_{ij}\ov{Q^C_i} \left(i\sigma^2\right) \sigma^I L_j S_3^I+\text{h.c.}\,,
    \label{eq:S1S3int}
\end{align}
\ie,~assuming lepton and baryon number conservation.\footnote{We also disregarded the scalar potential, which gives the $S_1+S_3$ mixing~\cite{Dorsner:2019itg,Crivellin:2020mjs,Crivellin:2020ukd} as the corresponding effect in meson physics are usually small.} Here, $Q$ ($L$) are quark (lepton) $SU(2)_L$ doublets, $u$ ($e$) are up-type quark (charged lepton) singlets, $\sigma^I~(I=1,2,3)$ are the Pauli matrices, and $i,j$ are flavor indices.

Since $R(D^{(*)})$ requires a tree-level effect for an explanation of the anomalies, this results via $SU(2)_L$ invariance, in general also in an effect of the same order in $b\to s\nu\ov\nu$. As it is here compared to the loop-suppressed SM effect, this excludes the possibility of (fully) accounting for data with $S_1$ or $S_3$ individually.\footnote{Note that while the $S_1$ LQ alone, even when supplemented with right-handed neutrinos~\cite{Chen:2025npb}, 
one cannot explain $R(D^*)$ and $b \to s \ell \ov{\ell}$.}
However, due to the relative sign difference in $b\to c\tau\ov\nu$ and $b\to s\nu\ov\nu$ in $S_1$ vs $S_3$, it is possible to have constructive interference in the former but destructive interference in the latter. In fact, we find that the following flavor structure corresponds to a minimal set of the couplings, including only non-zero values for third-generation leptons,
to explain the constructive interference in $b\to c\tau\ov\nu$ but the destructive interferences in both $b\to s\nu\ov\nu$  and $s\to d\nu\ov\nu$:
\begin{eqnarray}
  \lambda^L_{ij} &&=\lambda^L_{}\left(
  \begin{array}{ccc}
    0 &0 &r^3\\
    0 &0 &r\\
    0 &0 &1\\
  \end{array}
  \right)\,,
  ~~ \lambda^R_{ij} = \left(
  \begin{array}{ccc}
    0 &0 &0\\
    0 &0 &\lambda^R_{23}\\
    0 &0 &0\\
  \end{array}\right)\,,\notag\\
  \kappa_{ij} &&=(1-\epsilon)\lambda^L_{}\left(
  \begin{array}{ccc}
    0 &0 &(1-\delta)r^3\\
    0 &0 &-r\\
    0 &0 &1\\
  \end{array}
  \right)\,.
\label{eq:S1S3_coupling_structure}
\end{eqnarray}
Here, $\epsilon=0$ leads to an exact cancellation in $b\to s\nu\ov\nu$ (at tree-level for equal masses of $S_1$ and $S_3$).\footnote{Note that the misalignment between the $S_1$ and $S_3$ flavor structure could also be due to different masses and/or nontrivial mixing effects from the scalar potential.}
On the other hand, $\delta$ implies an additional degree of freedom for $\kappa_{13}$ coupling, resulting in a deviation from the CKM-like hierarchy encoded in $r\approx0.2$. In the following, we will call $\epsilon = \delta=0$ the alignment limit, where both contributions to $b \to s \nu \ov\nu$ and $s \to d \nu \ov\nu$ are exactly canceled at tree level. Note that thanks to this CKM-like alignment, leptoquark contributions to the $D$-meson mixing, $D_s^-\to \tau\ov\nu$,
and $B\to \pi\tau \ov\nu$ decays are suppressed and thus do not provide relevant constraints once the relevant uncertainties are taken into account. Finally, note that we are working in the down basis. This means that after electroweak symmetry breaking, the CKM matrix appears in the couplings of leptoquarks to the left-handed up-quarks.
 
\subsection{Singlet vector leptoquark model: \texorpdfstring{\boldmath{$U_1$}}{U1}}

References~\cite{Alonso:2015sja,Calibbi:2015kma,Fajfer:2015ycq} pointed out that $SU(2)_L$ singlet vector-leptoquark with hypercharge $ 2/3$, arising in the famous Pati-Salam model~\cite{Pati:1974yy}, has the potential to provide a combined explanation of the $B$ anomalies. Its phenomenology as a simplified model was studied in Refs.~\cite{Alonso:2015sja,Calibbi:2015kma,Fajfer:2015ycq,Hiller:2016kry,Bhattacharya:2016mcc,Buttazzo:2017ixm,Kumar:2018kmr} and Refs.~\cite{Barbieri:2015yvd,Barbieri:2016las,Assad:2017iib,DiLuzio:2017vat,Barbieri:2017tuq,Calibbi:2017qbu,Bordone:2017bld,Greljo:2018tuh,Bordone:2018nbg,Blanke:2018sro,Matsuzaki:2018jui,DiLuzio:2018zxy,Balaji:2018zna,Cornella:2019hct,Balaji:2019kwe,Fuentes-Martin:2020bnh,Guadagnoli:2020tlx,Dolan:2020doe,Iguro:2021kdw,King:2021jeo,Iguro:2022ozl,FernandezNavarro:2022gst,Fuentes-Martin:2020hvc,Crosas:2022quq} suggested different UV completions, as the original Pati-Salam model cannot be realized at the TeV scale due to the stringent constraints from $K_L\to\pi e\ov\mu$ etc. 

Here we work in a simplified model\footnote{Even though a theory with a massive vector boson without an explicit Higgs sector is not renormalizable, it has been shown that several loop effects in flavor observables can be consistently calculated, leading to numerically relevant effects which result in additional correlations~\cite{Crivellin:2018yvo}. Furthermore, it has been shown in Refs.~\cite{Fuentes-Martin:2020hvc,Cornella:2021sby} that the effects of heavy leptons, necessarily present in UV-complete models, are numerically small (for these observables).} with the quark-lepton interactions parametrized as
\begin{eqnarray}
\mathcal{L}_{U_1}=\left(\kappa_{ij}^{L}\overline{Q_i}\gamma_{\mu} L_{j}+\kappa_{ij}^{R}
\overline{d_i}\gamma_{\mu} e_{j}\right)U_{1}^{\mu}+\text{h.c.}\,.
\label{eq:U1int}
\end{eqnarray}
In addition to the notation introduced in Eq.~(\ref{eq:S1S3int}), $d$ is the $SU(2)_L$ singlet down-type quark. In the following, we will neglect the right-handed couplings $\kappa^R_{ij}$, which are unnecessary to explain the anomalies. We will focus on the situation in which the tauonic couplings are sizable while others are negligible, \ie, $\kappa_{i3}^{L}\neq 0$. Like for the $S_1+S_3$ model, we work in the down-basis.

\section{Phenomenology}
\label{sec:pheno}

We now perform an updated phenomenological analysis of the $S_1+S_3$ and the $U_1$ LQ models. In particular, we will investigate how we can distinguish the two models with the help of the neutrino modes $B\to K\nu\ov\nu$ and $K\to \pi\nu\ov\nu$ as these decays are sensitive to third-generation lepton interactions. We will use the formula for the various observables obtained in Refs.~\cite{Crivellin:2019dwb,Gherardi:2020qhc} for the $S_1+S_3$ model and Refs.~\cite{Capdevila:2017iqn,Crivellin:2018yvo,Cornella:2019hct,Iguro:2023rom} for the $U_1$ model. The interested reader is referred to these articles for details, such as the definitions of the loop functions etc.

\subsection{Observables and Input}

The SM predictions for several observables are sensitive to the numerical values of the CKM elements. In particular, $\mathcal{B}(K \to \pi\nu\ov\nu)$ is proportional to $A^4$, where $A$ is the parameter of the Wolfenstein parameterization mainly determining $|V_{cb}|$. In this study, we follow the strategy of Ref.~\cite{Allwicher:2024ncl} and obtain $|V_{cb}|$ from the global fit including both the inclusive~\cite{Finauri:2023kte} and the exclusive determinations~\cite{Martinelli:2023fwm,Bordone:2024weh} from semi-leptonic $b\to c\ell \ov\nu$ transitions, resulting in
\begin{align}
    |V_{cb}| = (41.37 \pm 0.81) \times 10^{-3}\,.
\end{align}
Note that the CKM parameters fitted by the UTfit collaboration~\cite{UTfit:2022hsi,Bonatalk},
 are obtained from $(\bar\rho, \bar\eta)$ without using the NP sensitive observables ($\varepsilon_k$, $B_{s,d}-\bar B_{s,d}$ mixing, and  semi-leptonic $B$ decays): 
\begin{align}
\lambda &= 0.2251 \pm 0.0008\,,\label{eq:lambda}\\
A&= 0.816 \pm 0.017\,,\\
\bar\rho & =  0.144 \pm 0.016\,,\\
\bar\eta & =  0.343 \pm 0.012\,.
\end{align}
Using these parameters, we obtain
\begin{align}
&|\lambda_t| \equiv |V_{tb} V^\ast_{ts}| =
0.0406 \pm 0.0009\,,\\
&|V_{ub}| = 0.00355\pm 0.00015\,.
\end{align}
The resulting SM predictions for $K^+ \to \pi^+ \nu\ov\nu$ and $B^+ \to K^+ \nu\ov\nu$ are given in the Introduction. Note that $R(D^{(\ast)})$ is independent of CKM elements but that the $b\to s\ell\ov\ell$ fit feebly depends on them via the total branching ratios while the angular observables, like $P_5^\prime$ are again independent.

Using the $|V_{ub}|$ value given above and the decay constant $f_{B^+} = 190.0(1.3)$\,MeV from lattice QCD (with $N_f= 2 + 1 + 1$)~\cite{FLAG2024}, we obtain~\cite{Zuo:2023dzn}
\begin{align}
    \mathcal{B}(B^-\to\tau\ov\nu)_{\text{SM}}=(0.80\pm0.06)\times 10^{-4}\,.
\end{align}
On the experimental side, including the recent Belle II result~\cite{Belle-II:2025ruy}, one has~\cite{Abitalk}
\begin{align}
\mathcal{B}(B^-\to\tau\ov\nu)=(1.12\pm 0.21)\times 10^{-4}\,.
\end{align}
While theory and experiment are consistent at $1.5\sigma$, the foreseen sharpening of the experimental data offers the possibility to unveil NP at the same level as suggested by $R(D^{(*)})$. Furthermore, for
\begin{align}
R(\pi)=\frac{\mathcal{B}\left(B\to\pi\tau\ov\nu\right)}{\mathcal{B}\left(B\to\pi\ell\ov\nu\right)}\,,
\end{align}
the SM prediction and experimental value are $R_\pi^{\rm SM}=0.719\pm0.028$ and $R_\pi^{\rm exp}=1.01\pm0.49$~\cite{Duan:2024ayo}. 
Due to the large experimental uncertainty, these results are not in statistically significant tension with the SM, but it is interesting to note that they lie above the SM prediction as well.

Concerning $b\to s\ell\ov{\ell}$ data, there is a plethora of observables that are included in the global fit which find large significances for new physics in $C_9^{\text{U}}$, above the 5$\sigma$ level~\cite{Alguero:2023jeh, Mahmoudi:2024rhb,Ciuchini:2022wbq,Gubernari:2022hxn} with best-fit values of around $C_9^{\text{U}}\approx -1$, \ie, a destructive effect of around 20\% w.r.t.~the SM. 
However, as $C_9^{\text{U}}$ could, at least in principle, be mimicked by (additional unknown) hadronic SM effects~\cite{Ciuchini:2022wbq}, we will not confine ourselves to a specific range but rather show contour lines for the NP contribution to $C_9^{\text{U}}$ in our plots. Note that both models predict the right sign in $C_9^{\text{U}}$, given a constructive effect in $R(D^{(*)})$.

\begin{figure*}[t]
{\includegraphics[width=0.65\textwidth]{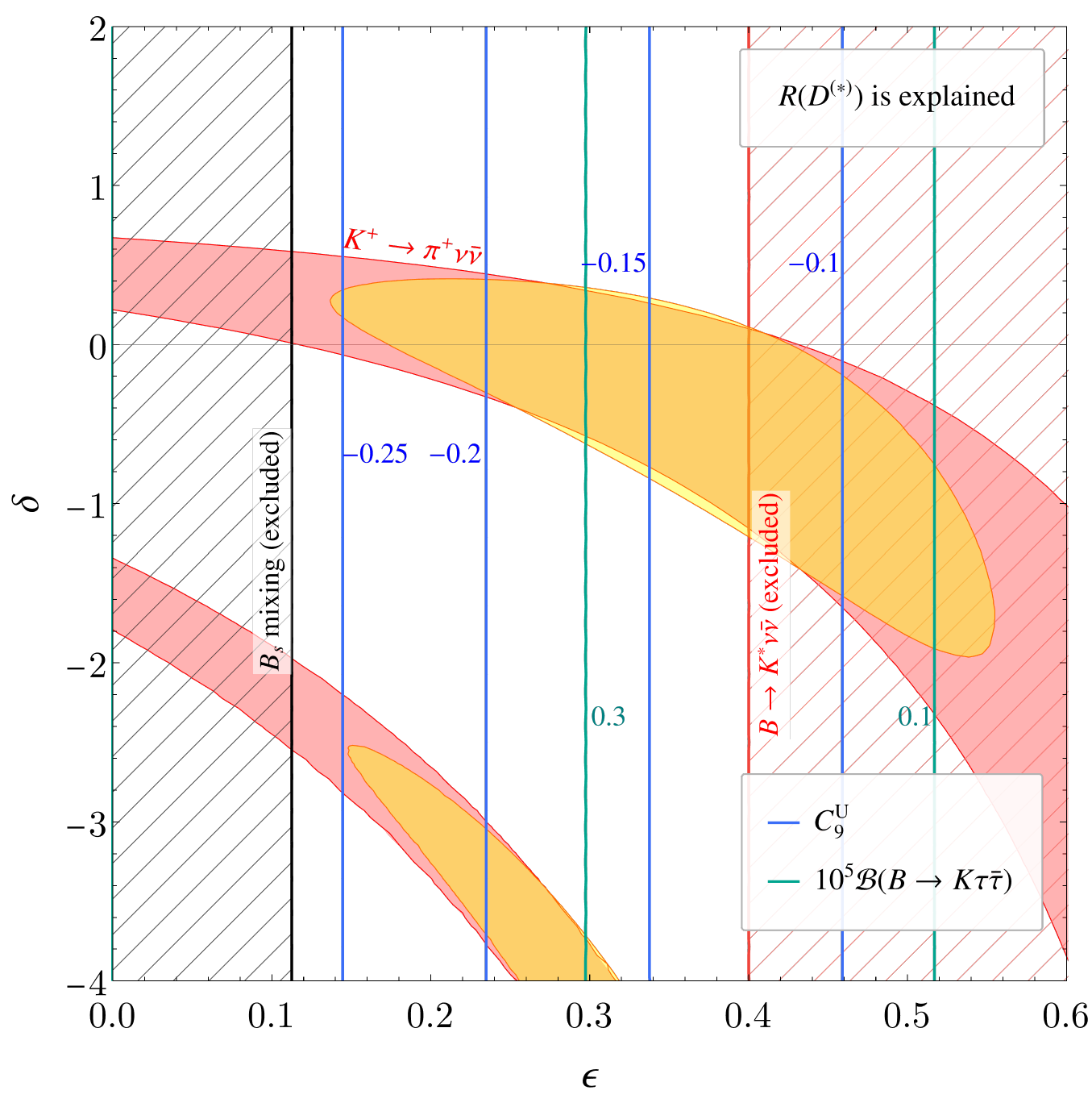}}
\caption{\label{fig:S1S3-2}
The preferred and excluded region in the $\epsilon$--$\delta$ plane for $m_{S_1} = m_{S_3} = 2\TeV$ and $\lambda^L =1, r = 0.2, \lambda^R_{23}= -0.5$  of the $S_1+S_3$ LQ model, such that entire shown plane can accommodate $R(D^{(\ast)})$ at the 1$\sigma$ level, \ie,~$R(D)\approx 0.35$ and $R(D^\ast)\approx 0.28$. 
The blue and green contours represent $C_9^{\text{U}}$ and Br$(B\to K\tau\ov\tau)\times 10^{-5}$, respectively. The red region is preferred by $K^+\to \pi^+ \nu\ov\nu$ ($1\sigma$), the yellow region is the global fit to $K^+\to \pi^+ \nu\ov\nu$ and $\mu(B^+ \to K^{+} \nu\ov\nu)$. $B_s-\bar B_s$ and $B\to K^*\nu\ov\nu$ exclude the black and red hatched regions, respectively. Note that the NP effect in $B^-\to \tau \ov\nu$ is small with our CKM-like coupling structure.}
\end{figure*}

In addition to the observables discussed in the introduction, we consider the constraints from the $\Delta F=2$ processes for the $S_1+S_3$ model. More specifically, we have effects in $B_s-\bar B_s$, $B_d-\bar B_d$ and the kaon mixing, which are correlated in a CKM-like fashion due to our choice of the hierarchical coupling structure. As $B_s-\bar B_s$ mixing results in the most stringent bounds in this setup, we do not show the other two cases and use the global fit result of Ref.~\cite{Bona:2022zhn}.\footnote{Note that the kaon mixing is however constraining for larger values of $r$ as can be seen in Fig.~\ref{fig:S1S3} the Appendix.} For our (simplified) $U_1$ model, the calculation of $\Delta F=2$ processes is plagued by divergences due to the lack of a Higgs mechanism generating its mass. As it has been shown that the bounds can be weakened in a UV-complete model by a GIM-like cancellation involving vector-like leptons, we opted not to show the bounds in the plots. However, it should be kept in mind that these limits cannot be fully avoided but are rather model-dependent.

In both the $S_1+S_3$ and the $U_1$ LQ model, we restrict our study to real couplings.
In fact, the $CP$-violating complex phases 
are not helpful to explain
 $b \to c \ell \overline{\nu}$ anomaly~\cite{Iguro:2018vqb}.
Furthermore, while for the $S_1+S_3$ model,  $\arg (r \lambda^L_{23} (\lambda^R_{23})^\ast)$ induces a non-zero electric dipole moment (EDM) of the tau lepton~\cite{Crivellin:2021spu} and a charm-quark (chromo) EDM at the one-loop level~\cite{Dekens:2018bci,Crivellin:2019qnh}, the bounds are weak.\footnote{The charm-quark tensor charge calculated with lattice QCD turns out to be small~\cite{Alexandrou:2019brg}. However, the charm EDM would contribute to the nucleon EDM, which can be observed in future experiments~\cite{Iguro:2023rom}. }
Moreover, the presence of $\arg r$ induces $CP$-violating contributions to $\Delta F=2$ processes, which could, however, also weaken the bounds. 

\subsection{\texorpdfstring{\boldmath{$S_1+S_3$}}{S1+S3} model}
\label{sec:pheno_S1S3}

For the $S_1+S_3$ model, we study the free parameters $\lambda^L$, $\lambda^R_{23}$, $\delta$, and $\epsilon$ for $r=0.2$ as motivated by the CKM hierarchy (and consent with kaon mixing), assuming all couplings are real, as discussed in the last section. Note that one can choose $\lambda^L>0$ without loss of generality.

First, for the alignment limit $\epsilon=\delta=0$, the effect in $ b \to s \nu \ov\nu$ transitions vanishes at tree level due to the cancellation between the $S_1$ and $S_3$ contributions, which comes from the coupling structure in Eq.~\eqref{eq:S1S3_coupling_structure}.\footnote{However, note that the effect in $B\to \pi\nu\ov\nu$ only vanishes for $\delta=2$.} In this limit, the discrepancy in $R(D^{(*)})$ can be explained via the combinations $\lambda^L_{33}\lambda^L_{23}$ and $\lambda^L_{33}\lambda^R_{23}$, the latter leading to scalar and tensor operators. However, the constraints from $B_s-\bar B_s$ mixing have to be respected. In fact, one can account for $R(D^{(*)})$ at the $1\sigma$ level without violating the bounds from $B_s-\bar B_s$ mixing for $m_1=m_3=2\,$TeV, $\lambda^L=1.0$ and $\lambda_{23}^R= - 0.5$ (see Fig.~\ref{fig:S1S3} in the Appendix for details). For these values, we show the dependence on $\delta$ and $\epsilon$ in Fig.~\ref{fig:S1S3-2}. Note that for $\epsilon\neq0$, NP affects both $\mu(K^+\to \pi^+ \nu\ov\nu)$ and $\mu(B^+\to K^+ \nu\ov\nu)$. 

Due to the $SU(2)_L$ invariance, $\lambda^L_{33} \lambda^L_{23}$ leads to a tree-level contribution to $b\to s \tau\ov\tau$~\cite{Capdevila:2017iqn} processes which results in a huge enhancement of them. However, even though the predictions are several orders of magnitude above the SM one, they are still significantly away from the current limits. Via the same combination of couplings, \ie,~$\lambda^L_{33} \lambda^L_{23}$, the model also generates $C_9^{\text{U}}$ via a tau-loop penguin with an off-shell photon. However, it is constrained from $\bbs$ mixing such that the current best-fit value for $C_9^{\text{U}}$ cannot be achieved. Nevertheless, it is important that the effect is not negligible and is predicted to have the correct sign (\ie,~negative) preferred by the global fit to $b\to s\ell\ov{\ell}$ data. 

The red band in Fig.~\ref{fig:S1S3-2} is consistent with the NA62 result at $1\sigma$ and the yellow region is obtained by combining this measurement with $B^+\to K^+\nu\ov\nu$, $R(D^{(\ast)})$ and $\mathcal{B}(B^- \to \tau \ov\nu)$. We find that the shift in $\mathcal{B}(B^-\to \tau\ov\nu)$ and $R(\pi)$ is small due to the partial cancellation between $S_1$ and $S_3$ contributions. The black-hatched region is excluded by $B_s-\bar B_s$ mixing. Besides, it is noted that the LHC constraint is imposed based on \texttt{HighPT} package~\cite{Allwicher:2022mcg} and found to be weak outside the plot range.\footnote{Note that $S_1$ could, in principle, give a chirally enhanced effect in $g-2$ of charged leptons via a top-quark loop~\cite{ColuccioLeskow:2016dox}. However, this can lead to dangerously large effects in $\tau\to\mu\gamma$, etc., if one aims at a combined effect in the $B$ anomalies~\cite{Crivellin:2019dwb}. Effects in EW oblique observables and Higgs signal strength originating from the scalar potential are mostly small, such that they can only be tested at future colliders~\cite{Crivellin:2020ukd}. We checked that the bounds from $\tau\to \mu\nu\ov\nu / \tau\to e\nu\ov\nu$ are satisfied.}

Finally, note the constraining power of the NA62 measurement. While the $B$ physics observables mainly depend on $\epsilon$, $\delta$ changes $K\to \pi \nu\ov\nu$ and is more relevant than the bounds from kaon mixing for small values of $r$ in the $S_1+S_3$ model.

\subsection{\texorpdfstring{\boldmath{$U_1$}}{U1} model}
\label{sec:pheno_U1}

\begin{figure*}[t]
\includegraphics[width=0.65\textwidth]{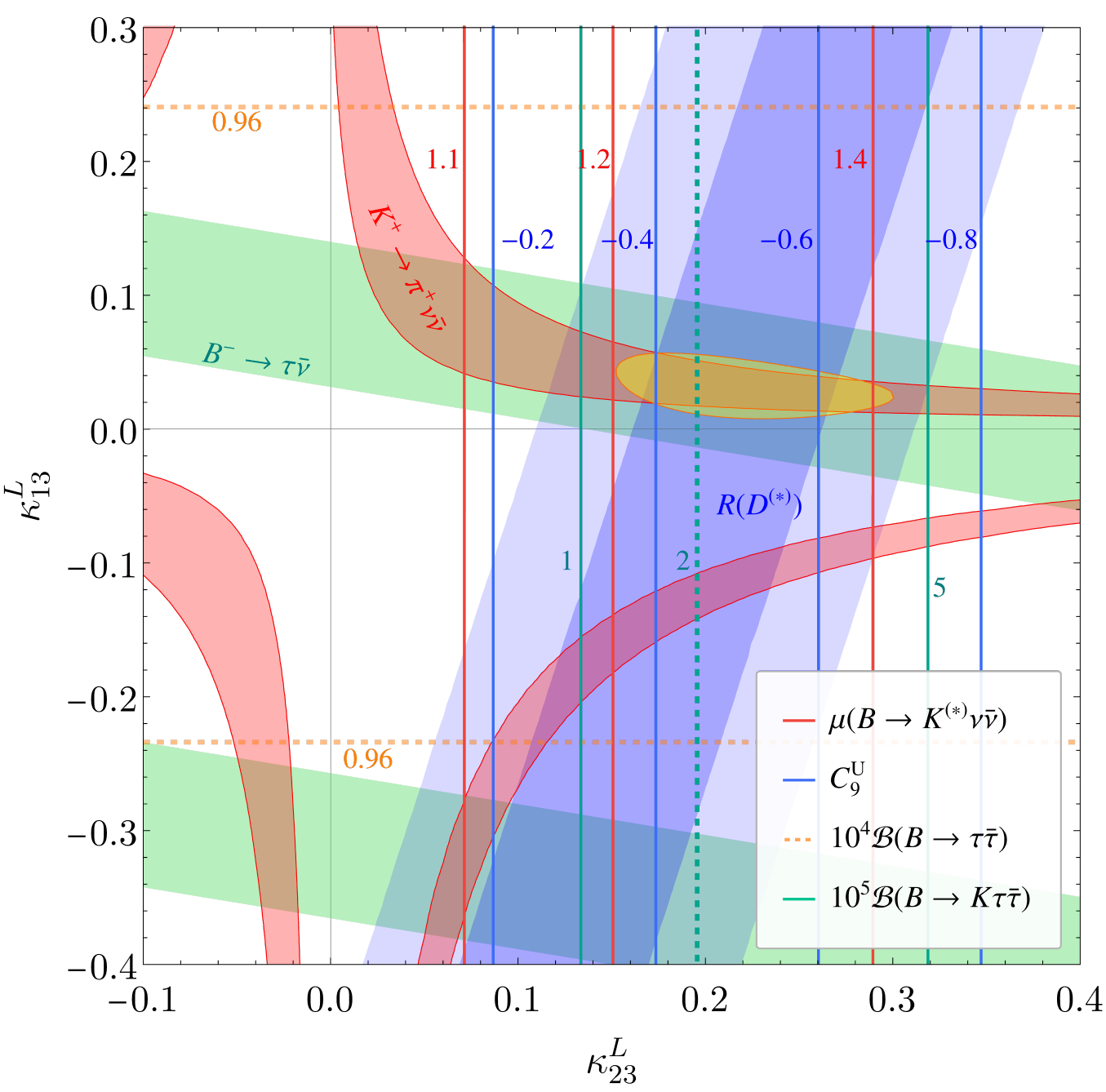}
\caption{\label{fig:U1}
For the $U_1$ LQ model we take $m_{U_1} = 2\TeV$ and $\kappa_L^{33}=1.5$. Then, the blue region can accommodate the $R(D^{(\ast)})$ (at 1$\sigma$, 2$\sigma$), while the red and green regions are consistent with $K^+ \to \pi^+ \nu \ov\nu$ and $B^- \to \tau\ov\nu$. The yellow area represents the global fit for these observables, including $B^+ \to K^+ \nu \ov\nu$. The contours are the predictions for $\mu(B^+ \to K^+ \nu\ov\nu)$, $C_9^{\text{U}}$, $\mathcal{B}(B \to \tau\ov\tau)$, and $\mathcal{B}(B \to K \tau\ov\tau)$ and the dashed lines show the Belle II sensitivity at $50\,\text{ab}^{-1}$\! of data \cite{Belle-II:2018jsg}.
}
\end{figure*}

For the $U_1$ LQ model we have less free relevant parameters and fix $\kappa_L^{33}=1.5$ and vary $\kappa_L^{23}$ and $\kappa_L^{13}$ in Fig.~\ref{fig:U1}. We show the regions preferred by $R(D^{(\ast)}),\,K^+\to \pi^+\nu\ov\nu$, and $ B\to\tau\ov\nu$ in blue, red, and green bands, respectively. For $R(D^{(\ast)})$, we show both the $1\sigma$ and $2\sigma$ regions, while for the other observables, only the $1\sigma$ regions are shown. The global fit based in $R(D^{(\ast)}),\,K^+\to \pi^+\nu\ov\nu, B^+ \to K^+ \nu\ov\nu$ and $B\to\tau\ov\nu$ is shown in yellow with $1\sigma$. The predictions for $C_9^{\text{U}}$,  $\mu(B\to K^{(\ast)}\nu\ov\nu)$, $\mathcal{B}(B\to\tau\ov\tau)$, and $\mathcal{B}(B\to K \tau\ov\tau)$ are shown in terms of contour lines.

In Fig.~\ref{fig:U1}, the orange and green dashed lines show the Belle II sensitivity for $\mathcal{B}(B\to\tau\ov\tau)$ and $\mathcal{B}(B\to K \tau\ov\tau)$ at $50\,\text{ab}^{-1}$\! of data \cite{Belle-II:2018jsg}. Interestingly, the yellow region, which is preferred by current data, can be probed by $\mathcal{B}(B\to K \tau\ov\tau)$, while the second favored region ($\kappa^L_{13} \approx -0.3$, which only consistent at the $2\sigma$ level) could be probed by $\mathcal{B}(B\to\tau\ov\tau)$. Note that $K\to\pi\nu\ov\nu$ provides a unique bound and constrains the product of $\kappa_L^{23}\times \kappa_L^{13}$ on the plane such that only one of these regions is favoured. 

In comparison to the $S_1+S_3$ model, only smaller effects in $B\to K\nu\ov\nu$ are possible in the $U_1$ model. Furthermore, $B^-\to \tau\ov\nu$ is more relevant and large absolute values of $C_9^{\text{U}}$ are possible. However, the latter depends crucially on the $B_s-\bar B_s$ mixing bound, which we ignored, as mentioned earlier, as it depends on the UV completion.

\section{Conclusion}
\label{sec:conc}

\begin{figure*}[t]
{\includegraphics[width=0.45\textwidth]{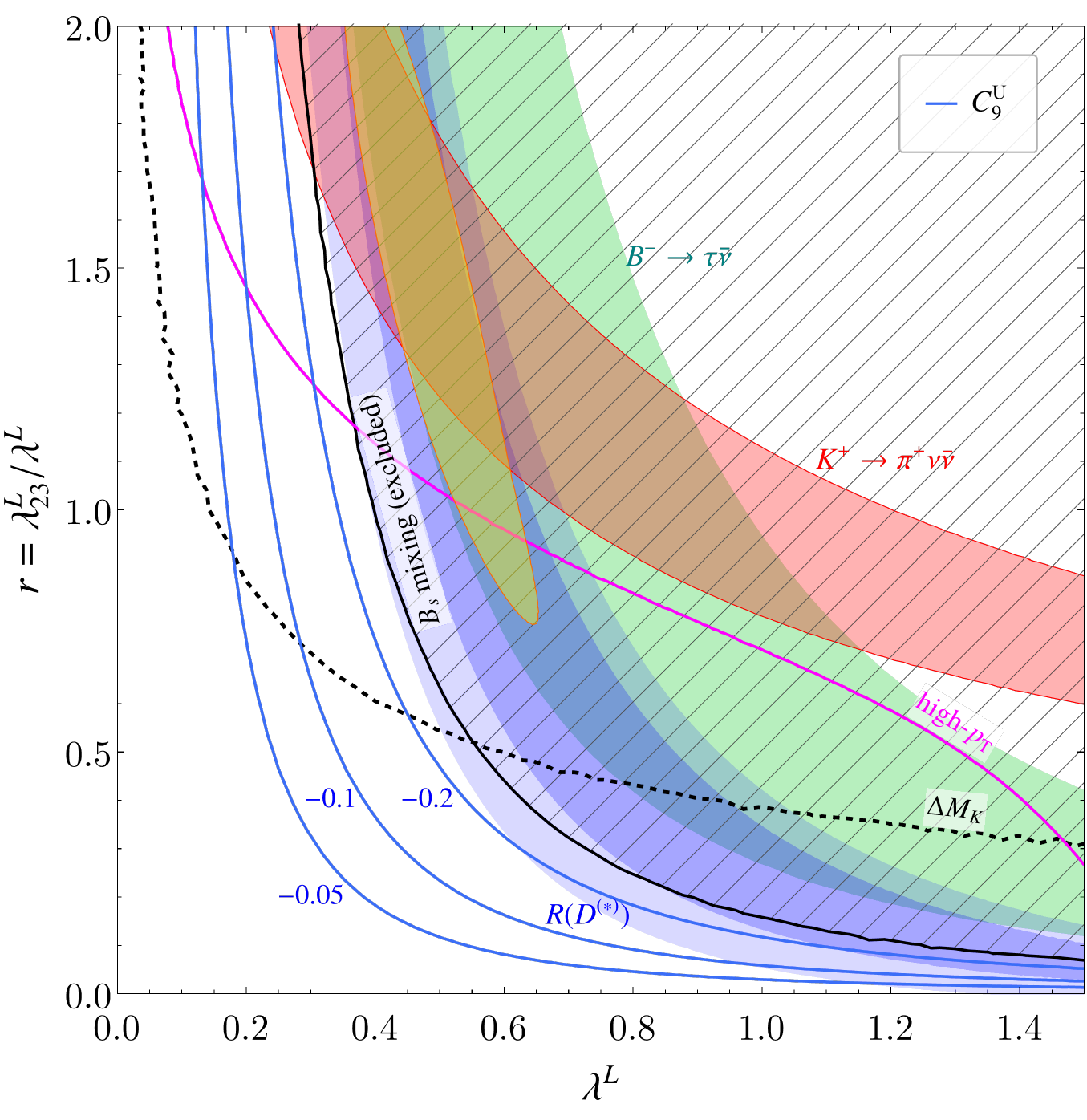}}\qquad
{\includegraphics[width=0.45\textwidth]{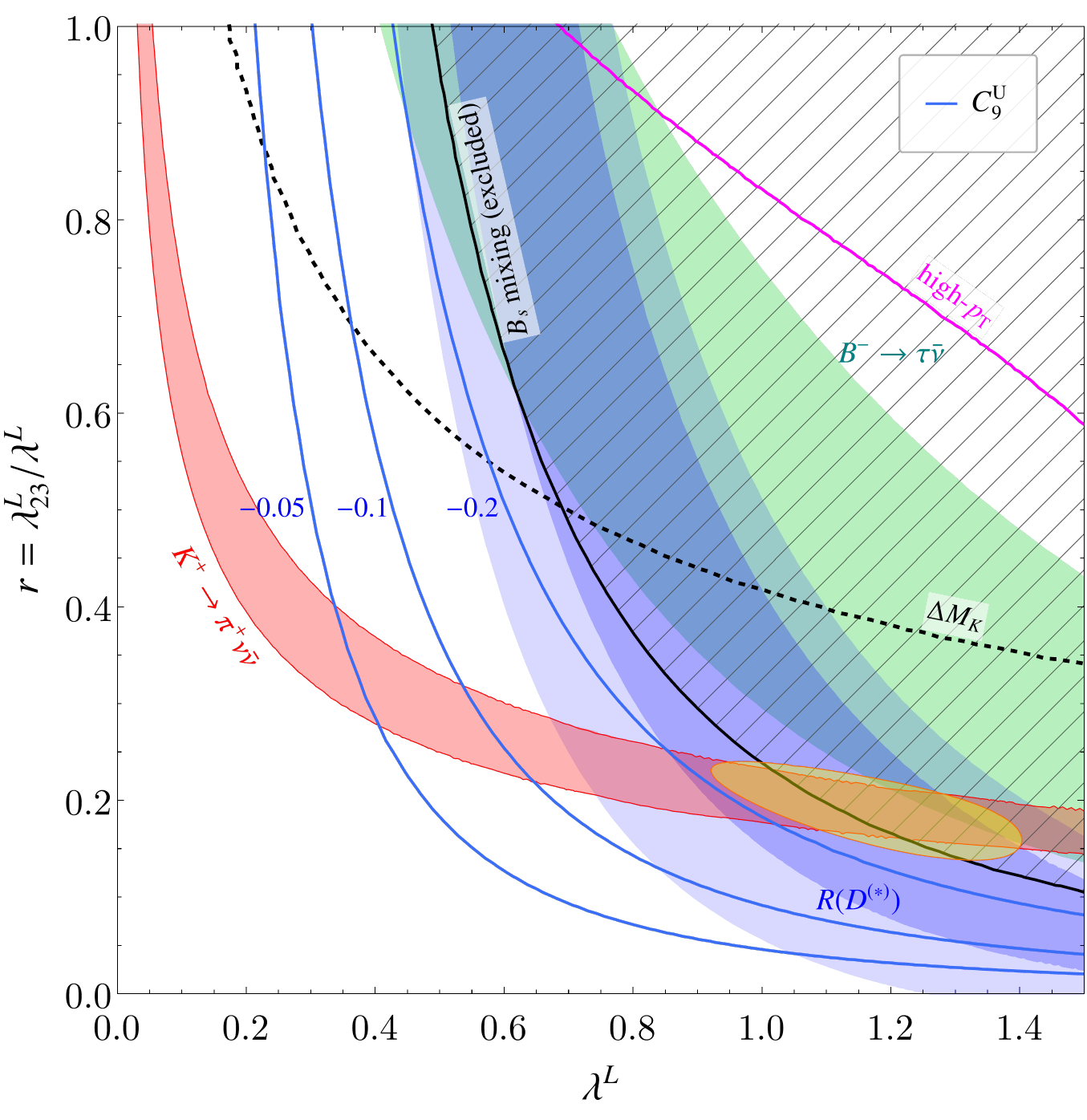}}
\caption{\label{fig:S1S3}
Preferred and excluded regions for the $S_1+S_3$ LQ model for the flavor structure given in Eq.~\eqref{eq:S1S3_coupling_structure}. We take a common mass $m_{S_1} = m_{S_3} = 2\TeV$ and $\lambda^R_{23}= -0.5$. In the left panel, we assume the alignment limit $\epsilon=\delta=0$, while on the right panel, we take $\epsilon=0.2$ but keep $\delta=0$. The blue region can accommodate the $R(D^{(\ast)})$ anomaly at the 1, 2$\sigma$ levels, while the red and green regions are consistent with the data of $K^+ \to \pi^+ \nu \ov\nu$ and $B^- \to \tau\ov\nu$. The yellow region represents the global fit for these observables (including $B^+ \to K^+ \nu \ov\nu$). The blue contours are the predictions of $C_9^{\text{U}}$.
The hatched region is excluded by the \bbms~bound at the $95\%$ CL, and the dashed lines represent the bound from the kaon mixing. Furthermore,  the upper-right regions of the magenta lines are excluded by the high-$p_{\text{T}}$ search~\cite{Allwicher:2022mcg}. 
 }
\end{figure*}

The confirmation of the $P_5^\prime$ and $R(D^{(*)})$ measurements together with the disappearance of the $R(K^{(*)})$ anomalies suggest that new physics might be only (or at least dominantly) directly related to tau leptons: A left-handed $\bar c b\bar \tau\nu$ operator affects the tau mode of $R(D^{(*)})$ and, via $SU(2)_L$ invariance, leads to an $\bar s b\bar \tau\tau$ operator which generates $C_9^{\text{U}}$ via an off-shell photon penguin. This setup can be achieved within two leptoquark models: the $SU(2)_L$ singlet-triplet scalar leptoquark ($S_1+S_3$) model and the (simplified) $SU(2)_L$ singlet vector leptoquark model $U_1$.

Because the mechanics for explaining these anomalies is the same in both models, the question arises how they can be discriminated. Here, the neutrino modes $b\to s\nu\ov\nu$ and $s\to d\nu\ov\nu$, which are both measured to be above the SM predictions (around a factor 2 larger with sizable errors) play a critical role as they involve tau neutrinos. In particular, kaon physics had been opening up the intensity frontier decades ago, and the latest NA62 result provided the first observation of the decay $K^+\to \pi^+ \nu\ov\nu$.

In particular, we find the following critical differences between the two models:
\begin{itemize}
\item While the achievable size of $C_9^{\text{U}}$ is smaller than the one suggested by the current global fits, 
it is important that the sign is predicted to be the preferred one. Note that for the same value of $R(D^{(*)})$ the $S_1+S_3$ model results in a smaller contribution to $C_9^{\text{U}}$ (and $b\to s \tau\ov\tau$ transitions) due to the presence of scalar and tensor amplitudes in $b\to c\tau \ov\nu$ transitions.
    \item The $U_1$ model only leads to a $\mathcal{O}(30\%)$ effects in $B \to K^{(\ast)}\nu\ov\nu$ as the effect is induced by a $W$ loop. On the other hand,
    for the $S_1+S_3$ model, where these decays receive tree-level leptoquark contributions,
    $\mathcal{O}(100\%)$ effects 
    can be easily obtained when the $S_1+S_3$ alignment is slightly broken.
\item While an enhancement in $b\to s \tau\ov\tau$ is observed in the $S_1+S_3$ scenario, the one obtained in the $U_1$ model is, in general, larger due to the purely left-handed vector effect in $R(D^{(\ast)})$. Thus, Belle II can probe the $U_1$ leptoquark model only in the $B \to K \tau \ov \tau$ and $B\to\tau\ov\tau$ channels.
        \item {In $B\to\tau\ov\nu$ and $B\to\pi\tau\ov\nu$, the $S_1+S_3$ scenario predicts relatively small effect due to the constraints on $r$ from the the kaon mixing. On the other hand, the $U_1$ scenario can lead to an $\mathcal{O}(30\%)$ shift w.r.t.~the SM when the model explains $R(D^{(\ast)})$ and the recent NA62 result. However, including  $\lambda^R_{13}$ as a free parameter, also $b\to u\tau\ov\nu$ transition can be significantly enhanced due to the scalar operator.}
\end{itemize}

\vspace{5pt}
{\it Acknowledgments} --- {\small 
We would like to thank Takeru Uchiyama for inspiring discussions. A.C. is supported by a professorship grant from the Swiss National Science Foundation (No.\,PP00P2$\_$211002). This work is supported by the JSPS Grant-in-Aid for Scientific Research Grant No.\,22K21347 (S.I.), 24K22879 (S.I.), 24K22872 (T.K.), 25K07276 (T.K.), and 25K17385 (S.I.). This work is also supported by JSPS Core-to-Core Program Grant No.\,JPJSCCA20200002. The work of S.I. is also supported by the Toyoaki Scholarship Foundation.
}

\appendix
\section{Additional analysis for \texorpdfstring{\boldmath{$S_1+S_3$}}{S1+S3} model}\label{sec:app}

In Fig.~\ref{fig:S1S3}, 
we show the preferred and excluded regions in the $\lambda^L$--\,$r$ plane, where $\lambda^R_{23}$ is fixed to be $-0.5$. In the left panel, the alignment limit $\epsilon = \delta =0$ is taken, while it is slightly broken ($\epsilon=0.2$) in the right panel. The darker and lighter blue regions can accommodate the $R(D^{(\ast)})$ anomaly within $1$ and $2\sigma$ levels. 
The red and green bands are consistent with the NA62 result and $\mathcal{B}(B^- \to \tau \ov\nu)$ at $1\sigma$ level, respectively.
The global fit based on $R(D^{(\ast)})$, $K^+\to\pi^+\nu\ov\nu$, $B^+\to K^+\nu\ov\nu$ and $B^- \to \tau\ov\nu$ is shown in the yellow ellipse. The blue contours represent the predictions of $C_9^{U}$. The hatched regions are excluded by $B_s-\bar B_s$ mixing, and bounds from the kaon mixing are shown by the dashed lines, which give a complementary constraint that is dominant for large values of $r$.  Furthermore, it is noted that the LHC constraint, shown by the magenta lines, based on \texttt{HighPT} package~\cite{Allwicher:2022mcg} is found to be weak compared to the $\Delta F=2$ bounds.

It is clearly shown that the alignment limit does not allow for a full explanation of $R(D^{(\ast)})$ due to the $\Delta F=2$ bounds, while a slight violation of this limit ($\epsilon=0.2 $) is enough to be consistent within the $1\sigma$ level, giving rise to effects in the neutrino modes. Thus, a simultaneous explanation of $R(D^{(\ast)})$, $\mu(K^+\to\pi^+\nu\ov\nu)$ and $\mu(B^+\to K^+\nu\ov\nu)$ is possible in the yellow region in the right panel with the prediction of $C_9^{U} \approx -0.2$. We found that the shift in $\mathcal{B}(B^-\to \tau\ov\nu)$ (and also $R(\pi)$) is negligible due to the partial cancellation between the $S_1$ and $S_3$ contributions. 

\bibliographystyle{utphys28mod}
\bibliography{VLQ}

\end{document}